\documentclass[aps,showkeys,noeprint,superscriptaddress,preprint]{revtex4-1}
\usepackage{graphicx}
\usepackage{braket}
\usepackage{amsmath}
\usepackage{bm}
\usepackage[british]{babel}
\graphicspath{{../Figures/}}

\begin{document}

\title{Direction-Dependent Parity-Time Phase Transition and Non-Reciprocal Directional Amplification with Dynamic Gain-Loss Modulation}

\author{Alex Y.\ Song}
\affiliation{Department of Electrical Engineering, Stanford University, Stanford, CA94305}
\author{Yu Shi}
\affiliation{Department of Electrical Engineering, Stanford University, Stanford, CA94305}
\author{Qian Lin}
\affiliation{Department of Applied Physics, Stanford University, Stanford, California 94305}
\author{Shanhui Fan}
\email[]{shanhui@stanford.edu}
\affiliation{Department of Electrical Engineering, Stanford University, Stanford, CA94305}

\date{\today}
\let\oldDelta\Delta
\renewcommand{\Delta}{\text{\scalebox{0.75}[1.0]{$\oldDelta$}}}

\begin{abstract}
    We show that a dynamic gain-loss modulation in an optical structure can lead to a direction-dependent parity-time ($\mathcal{PT}$) phase transition. The phase transition can be made thresholdless in the forward direction, and yet remains with a non-zero threshold in the backward direction. As a result, non-reciprocal directional amplification can be realized.

\end{abstract}


\maketitle

There have been significant recent interests in the fundamental quantum physics related to parity-time ($\mathcal{PT}$) symmetry, as well as in the applications of $\mathcal{PT}$ symmetry in both optical and electromagnetic structures \cite{Bender1998,Makris2008,Guo2009,Ruter2010,Lin2011,Chong2011,Bittner2012,Feng2013,Lumer2013,Feng2014,Hodaei2014,Cerjan2016}.
In particular, the connection between $\mathcal{PT}$ symmetry and non-reciprocity has been extensively discussed \cite{Huang2017,Yin2013,Zhou2016,Ramezani2010,Peng2014,Chang2014,Nazari2014,Zhu2014,Longhi2009}.
Optical structures exhibiting $\mathcal{PT}$ symmetry are typically described by scalar, time-independent dielectric functions. These structures cannot exhibit any non-reciprocity in their linear optical properties \cite{Jalas2013,Huang2017,Yin2013,Fan2012}.
In order to achieve non-reciprocal response, most existing works exploit the significant nonlinearity-enhancement provided by the $\mathcal{PT}$-phase transition \cite{Huang2017,Zhou2016,Ramezani2010,Peng2014,Chang2014,Nazari2014,Zhu2014}. Nevertheless, it has been shown that nonlinear non-reciprocal devices are fundamentally constrained by dynamic reciprocity, which significantly limits the practical functionalities of these devices \cite{Shi2015}.

In this paper we propose an alternative route to achieve non-reciprocity in $\mathcal{PT}$-symmetric structures.
We show that non-reciprocal directional amplification can arise in  structures under a dynamic material gain-loss modulation.
In particular, the gain-loss modulation induces a direction-dependent $\mathcal{PT}$ phase transition that is  thresholdless in the forward direction \cite{Ge2014,Feng2013,Cerjan2016}, but with a non-zero threshold in the backward direction. Consequently, such a structure enables direction-dependent amplification.

Related to our work, it has been shown that the dynamic modulation of the real part of the dielectric constant can be used to construct optical isolators and circulators \cite{Yu2009,Wang2013,Kang2011,Lira2012}. The underlying dynamics in the systems in Ref.~\onlinecite{Yu2009,Wang2013,Kang2011,Lira2012} however is Hermitian and is qualitatively different from the non-Hermitian physics that we discuss here, which arise from the modulation of the imaginary part of the dielectric constant.
Non-reciprocal directional amplification has been theoretically considered in Ref.~\onlinecite{Metelmann2015,Malz2018,Kamal2017,Ruesink2016,Abdo2013,Koutserimpas2018}, and experimentally implemented using Josephson junctions or optomechanical interactions \cite{Abdo2013,Fang2017,Shen2016,Ruesink2016}.
None of these works however made use of $\mathcal{PT}$-symmetry concepts.
Our approach points to a previously unrecognized connection between $\mathcal{PT}$-symmetry and non-reciprocal physics.
From a practical point of view, unlike all existing approach to directional amplification, the proposed scheme here does not rely upon the use of resonators and is inherently broad-band.
Furthermore, the gain-loss modulation is more straightforwardly integrable with standard semiconductor laser structures, and can be employed to protect laser sources from back-propagating noises.

\begin{figure}[b]
  \centering
  \includegraphics[width=246pt]{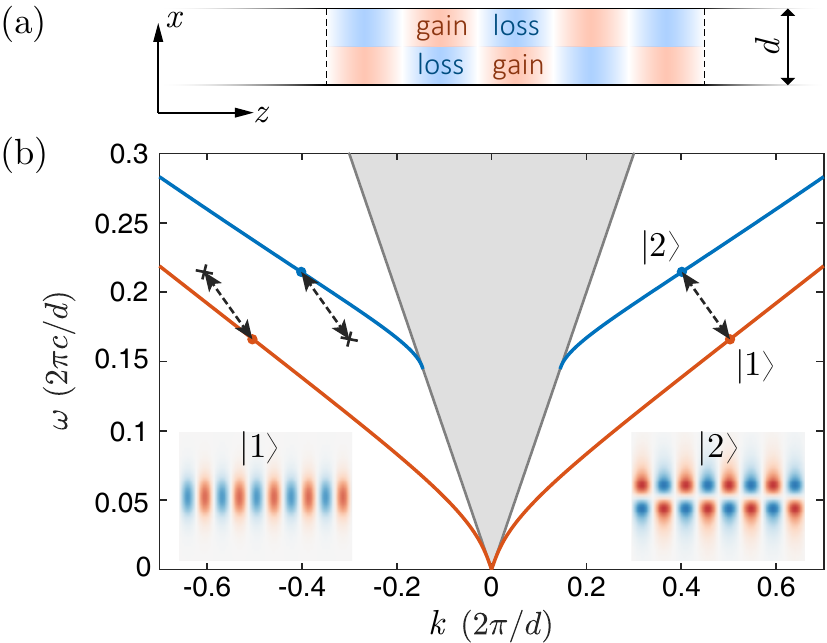}\\
  \caption{
    (a) Schematic of a dielectric waveguide under gain-loss modulation. The waveguide has a width of $d$. Gain-loss modulation is applied in the region indicated between the dashed lines.
    (b) Band structure of the dielectric waveguide. The red and blue curves show the even and the odd bands. The light cone is indicated by the gray shaded region. Modes $\ket{1}$ and $\ket{2}$ in the forward ($k>0$) propagating direction are coupled by the gain-loss modulation. Insets: the electric field modal profile of modes $\ket{1}$ and $\ket{2}$, respectively.
    }
  \label{fig:waveguide_modes}
\end{figure}

To illustrate the basic concept, we consider the dielectric waveguide structure schematically  shown in Fig.~\ref{fig:waveguide_modes}a.
The gain and loss in the waveguide is modulated as a function of space  and time.
Mathematically, we can represent the modulation in gain and loss by a time varying conductivity as
\begin{equation}
    \label{eq:modulation_profile}
    \tilde{\sigma}(x,z, t) =  \delta\sigma \, f(x) \cos(qz-\Omega t + \phi)
\end{equation}
Here, $\delta\sigma$ is the modulation strength.  $f(x)$ is the modulation profile in the $x$ direction.
$q$ is the wavevector. $\Omega$ is the modulation frequency. $\phi$ is the modulation phase.
We assume $f(x)$ is an odd function of $x$.
In an laser waveguide, gain and loss modulations can be achieved by controlling the pumping levels at different positions.

The waveguide without modulation has a photonic band structure as shown in Fig.~\ref{fig:waveguide_modes}b, which has two bands of modes that are even or odd with respect to the center plane of the waveguide. The field profiles of these modes are shown in the inset of Fig.~\ref{fig:waveguide_modes}b. Here for simplicity we consider only transverse-electric modes, which have the electric field perpendicular to the $xz$-plane.
In general, the modulation profile in Eq.~\ref{eq:modulation_profile} can couple modes from the two bands with opposite symmetry, with their frequencies separated by $\Omega$.
Considering only two modes involved in the coupling, the electric field in the waveguide can be written as
\begin{equation}
    E(x,z,t) = a_1\ket{1}+a_2\ket{2},\ \ \ket{1,2} = E_{1,2}(x) e^{i(k_{1,2}z-\omega_{1,2} t)}
\end{equation}
where $E_{1,2}(x)$ is the modal profile in $x$, which are normalized so that ${\lvert a_{1,2} \rvert}^2$ are the intensity of the respective modes. $k$ and $\omega$ are the wavevector and the frequency of each mode.
Defining $\psi(z) = (a_1(z), a_2(z))^T$, the equation of motion in  the modulated waveguide can be derived using coupled mode theory \cite{Yu2009}:

\begin{equation}
    \label{eq:eh}
    \begin{aligned}
        i\partial_z
        \psi(z)
        &=
        H(z)
        \psi(z),\\
        H(z) &=
        \left(\begin{array}{cc}
            0   &   -iCe^{-i\Delta kz-i\phi}\\
            -iCe^{i\Delta kz+i\phi} & 0
        \end{array}\right)
    \end{aligned}
\end{equation}
where $\Delta k = k_1 - k_2 - q$ is the wavevector mismatch, and $C=\frac{\delta \sigma}{8}\int{f(x)E_1(x)E_2(x)dx}$ is the coupling strength.

Eq.~\ref{eq:eh} is in the form of a time-periodic Schr\"{o}dinger equation with $z$ taking the role of time.
The Hamiltonian $H(z)$ satisfies the $\mathcal{PT}$ symmetry defined as \cite{Moiseyev2011,Luo2013,Joglekar2014,Chitsazi2017}
\begin{equation}
    \label{eq:PT}
    \mathcal{P}= \left(\begin{array}{cc}
    1   &   0   \\
    0   &   -1  \\
    \end{array}\right);
    \quad\quad
    \mathcal{T} H(z) \mathcal{T}^{-1} = H^*(-z)
\end{equation}
The definition of $\mathcal{P}$ stems from the fact that mode $\ket{1}$ is even  under parity operation, while mode $\ket{2}$ is odd.

As a result of the $\mathcal{PT}$ symmetry, the Floquet quasi-energies of the system must be either real or complex conjugate pairs \footnote{Please refer to the supplementary for more information.}.
To obtain the Floquet eigenstates and the quasi-energies, we first solve for the evolution operator $U(z,0)$ defined by $\psi(z)=U(z,0)\psi(0)$:
\begin{widetext}
\begin{equation}
    \label{eq:tm}
    U(z,0)  =
    \left(
            \begin{array}{cc}
                e^{i {\Delta} k z/2} \left(\cosh C' z -  i\frac{{\Delta} k/2}{C'} \sinh{C' z}\right)         &   -e^{-i\phi}e^{-i\Delta k z/2}\frac{C}{C'} \sinh{C'z} \\
                -e^{i\phi}e^{i\Delta k z/2} \frac{C}{C'} \sinh{C'z}    &   e^{-i\Delta k z/2} \left(\cosh{C'z} + i\frac{\Delta k /2}{C'}\sinh{C'z}\right)
            \end{array}
    \right)\\
\end{equation}
\end{widetext}
where $C' = \sqrt{C^2 - (\Delta k/2)^2}$. Then, the quasi-energy $\epsilon$ of the system can be obtained by letting $e^{-i\epsilon \zeta}\psi(0)=U(\zeta,0)\psi(0)$, where $\zeta=2\pi/{\Delta k}$ is the period of the Hamiltonian along $z$. The obtained quasi-energies are
\begin{equation}
    \label{eq:qe}
    \epsilon_{\pm} = \frac{\Delta k}{2}\pm C\sqrt{(\frac{\Delta k}{2C})^2-1}\ \pmod{\Delta k}
\end{equation}

\begin{figure}[tb]
  \centering
  \includegraphics[width=246pt]{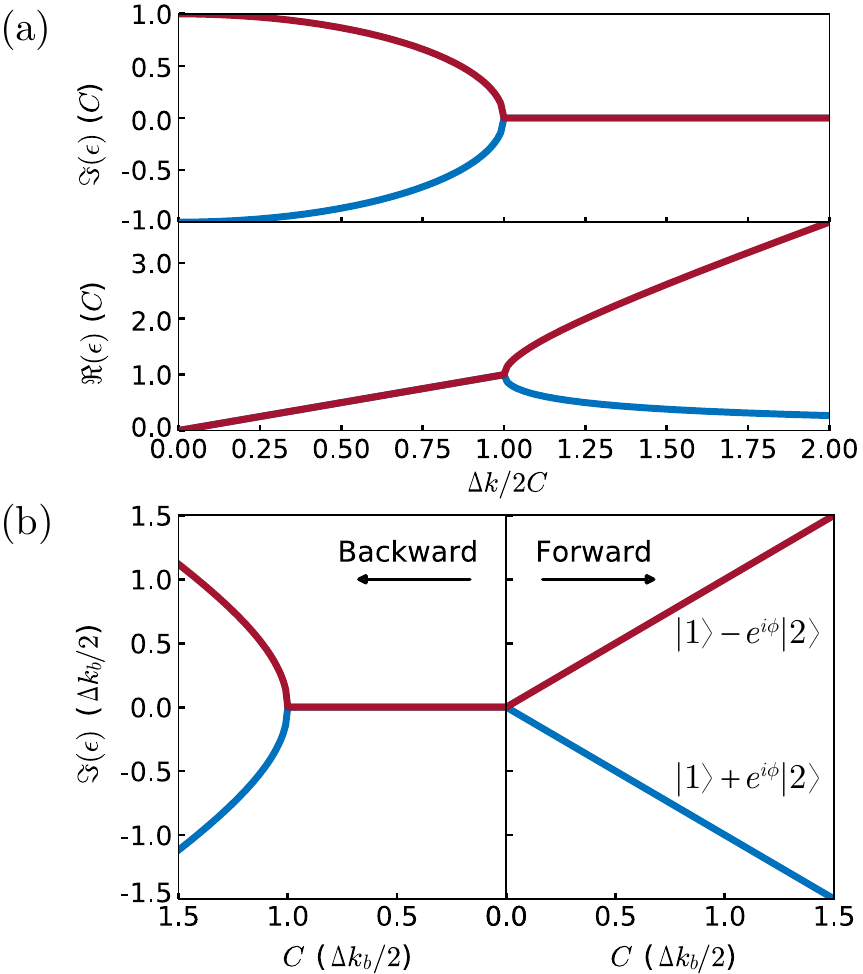}\\
  \caption{
    (a) Real and imaginary parts of the Floquet quasi-energies as a function of $\protect\Delta k/2C$. For large phase mismatch ($\protect\Delta k/2C>1$), the system is in the
    exact $\mathcal{PT}$ phase where the quasi-energies are real. For small phase mismatch ($\protect\Delta k/2C<1$), the system is in the broken phase.
    (b) Direction-dependent $\mathcal{PT}$ phase transition for the structure shown in Fig.~\ref{fig:waveguide_modes}. For the backward direction, due to strong phase-mismatch, the system is in the exact $\mathcal{PT}$ phase for modulation strength $C\leq \protect\Delta k_b/2$. For forward direction where $\protect\Delta k=0$,
    the system is in $\mathcal{PT}$ broken phase for any non-zero modulation strength.
    }
  \label{fig:PT_pt}
\end{figure}

From Eq.~\ref{eq:qe}, we observe that the system has a $\mathcal{PT}$ phase transition controlled by the ratio $\Delta k/2C$ as shown in Fig.~\ref{fig:PT_pt}a.
If the wavevector mismatch dominates over the coupling strength, i.e. $\Delta k > 2C$, the system is in the exact $\mathcal{PT}$ phase. Both quasi-energies are real, and the Floquet eigenmodes do not experience gain or loss.
On the other hand, for small wavevector mismatch, i.e. $\Delta k < 2C$ the system is in the broken phase where the quasi-energies of the system split into complex conjugate pairs. Thus, one of the Floquet modes will be amplified in the system.

We now consider a gain-loss modulation that provides a phase-matched coupling between two modes with different wavefectors in the forward direction as illustrated in Fig.~\ref{fig:waveguide_modes}b.
Such a modulation introduces a direction-dependent $\mathcal{PT}$ phase transition as is shown in Fig.~\ref{fig:PT_pt}b.
In the forward direction, $\Delta k_f=k_1-k_2-q=0$. From Eq.~\ref{eq:qe}, the quasi-energies become $\epsilon_{\pm}=\pm i C$. Thus the quasi-energies split into complex conjugate pairs as soon as coupling strength $C$ increases from $0$, i.e. the system exhibits a thresholdless $\mathcal{PT}$ phase transition \cite{Ge2014,Feng2014,Cerjan2016}.
For the backward direction however, $\Delta k_b \ne 0$, and the gain-loss modulation does not provide phase-matched coupling between any pair of modes.
Thus, the system  exhibits a $\mathcal{PT}$ phase transition with a non-zero threshold in the backward direction.

Such a direction-dependent $\mathcal{PT}$ phase transition gives rise to the effect of non-reciprocal directional amplification in this system. To illustrate this effect, we consider the regime where $\Delta k_f = 0$, and $\Delta k_b \gg C$. Under these conditions, the evolution operator can be simplified as
\begin{equation}
    \label{eq:tm_pm}
    U_f =
    \left(
        \begin{array}{cc}
            \cosh{C z}      &   -e^{-i\phi}\sinh{C z} \\
            -e^{i\phi}\sinh{C z}      &   \cosh{C z}
        \end{array}
    \right),\quad
    U_b =
    \left(
        \begin{array}{cc}
            1      &   0 \\
            0     &  1
        \end{array}
    \right)
\end{equation}
where ${f,b}$ stand for forward and backward propagations, respectively.
In the forward direction, one of the Floquet eigenmode, $\ket{1}-e^{i\phi}\ket{2}$, is amplified, while the other mode, $\ket{1}+e^{i\phi}\ket{2}$, is attenuated.
Thus, if $\ket{1}$  is the input to the waveguide, then the output becomes  $\cosh (Cz)\ket{1} - e^{i\phi}\sinh (Cz)\ket{2}$, providing amplification to the input mode.
In fact, input in the forward direction with any modal profile including $\ket{1}$, $\ket{2}$ or any of their combination except $\ket{1}+e^{i\phi}\ket{2}$, will be amplified.
In contrast, the system is in the exact $\mathcal{PT}$ phase in the backward direction. The Floquet eigenmodes are $\ket{1}$ and $\ket{2}$, with both quasi-energies approaching 0.
Thus an input mode in the backward direction, with any modal profile, does not experience any gain or loss.

We notice that in general  the system described by Eq.~\ref{eq:eh} is non-Hermitian, and the evolution operator $U$ is not unitary.
Thus, in general mode propagation in the system does not preserve mode orthogonality or the total energy flux.
Instead, for any $\Delta k$, the evolution operator $U$ in Eq.~\ref{eq:tm} is sympletic satisfying $\det(U)=1$. Hence, with any input mode profile, the intensity difference between modes $\ket{1}$ and $\ket{2}$ is always conserved.
This conservation law is different from that for a system undergoing dynamic modulation in the real part of the dielectric function \cite{Yu2009}, where the sum of the total photon number flux is conserved.

\begin{figure}[t]
  \centering
  \includegraphics[width=246pt]{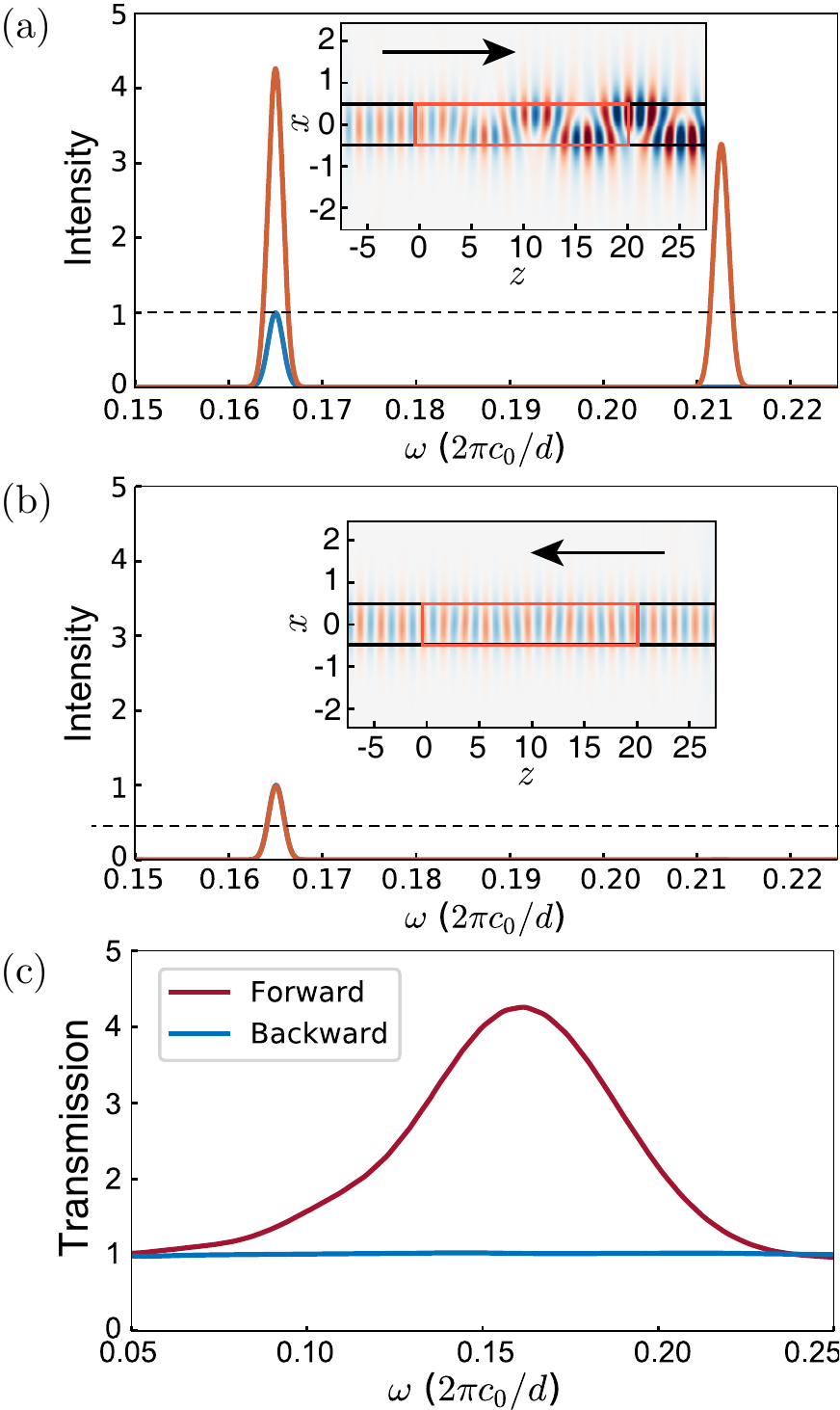}\\
  \caption{
    (a) and (b), transmission of an input pulse in mode $\ket{1}$ in the forward and backward direction, respectively.
    The intensity spectrum of the input pulse is shown by the blue curve.
    The output spectra are shown by the red curves.
    Electric field distribution are shown in the insets of (a) and (b), in which the wave propagation direction is marked by the black arrow.
    The gain-loss modulated region is indicated in the red rectangle.
    (c) The mode-to-mode transmission coefficient spectrum from mode $\ket{1}$ to mode $\ket{1}$ in the forward (red) and backward (blue) directions, respectively.
    }
  \label{fig:m1in_dg_fb}
\end{figure}

In the following, we numerically demonstrate the non-reciprocal effects predicted above using  finite-difference time-domain (FDTD) simulations.
We assume a waveguide with a permittivity of $\varepsilon=12.75$, and a width of $d=1$.
We select two modes of the waveguide as shown in Fig.~\ref{fig:waveguide_modes}b. Mode $\ket{1}$ has a frequency of $\omega_1 = 0.165$ and a wavevector of $k_1=0.5$, while mode $\ket{2}$ has  $\omega_2 = 0.213$ and $k_2=0.4$. The frequencies and the wavevectors are normalized to $2\pi c_0/d$ and $2\pi/d$, respectively. All the lengths below are normalized to $d$.
A section of the waveguide with a length of $l=20$ is under gain-loss modulation.
The modulation has a frequency $\Omega=0.048$ and a wavevector $q=0.1$, chosen to match the two modes in the forward direction.
The modulation strength $\delta \sigma $ is 1.
We input a Gaussian pulse in mode $\ket{1}$ from either left or right with a normalized peak intensity of 1 as shown in Fig.~\ref{fig:m1in_dg_fb}a.
In the forward direction, the input mode $\ket{1}$ evolves into a linear superposition of $\ket{1}$ and $\ket{2}$, as is shown in Fig.~\ref{fig:m1in_dg_fb}a.
The intensity in both modes $\ket{1}$ and $\ket{2}$ exceed unity, indicating the presence of amplification. The intensity difference between the modes $\ket{1}$ and $\ket{2}$ remains unity, in agreement with the conservation law derived analytically above.
In the backward direction however, the input mode $\ket{1}$ passes through without amplification or attenuation as is shown in Fig.~\ref{fig:m1in_dg_fb}b, again in agreement with the analytical results derived above. The numerical simulation here thus provides a validation of the theory presented above.

The non-reciprocal directional amplification discussed here can operate over a broad bandwidth, provided that modes $\ket{1}$ and $\ket{2}$ are in the parallel region of the even and odd bands in Fig.~\ref{fig:waveguide_modes}b.
Then, if a gain-loss modulation induces a phase-matched coupling between two modes with $\omega_1, k_1$ and $\omega_2, k_2$, it also induces a phase-matched coupling between modes $\omega_1+\delta\omega, k_1+\delta k$ and $\omega_2+\delta\omega, k_2+\delta k$ \cite{Yu2009}.
As a demonstration, in Fig.~\ref{fig:m1in_dg_fb}c we show the the mode-to-mode transmission spectrum for the even mode $\ket{1}$ in both directions.
In the forward direction, the transmission exceeds unity in a broad frequency range of $0.06$-$0.24$, while in the backward direction the transmission is nearly constant at $1$.
Thus, significant non-reciprocal directional gain can occur in a broad frequency range with its width comparable to its center frequency.
As a result, in practical device applications, the operation bandwidth will only be limited by the gain bandwidth of the materials.
The broadband characteristics here is in contrast with existing schemes on directional amplifications, which are all based on resonant interactions and hence are inherently narrow-banded.

\begin{figure}[t]
  \centering
  \includegraphics[width=245pt]{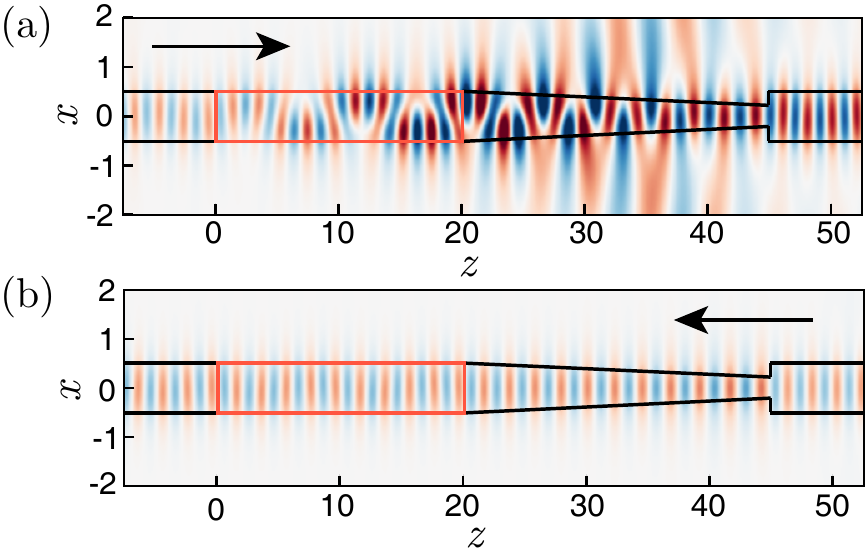}\\
  \caption{
    A visualization of the directional gain in a waveguide with gain-loss modulation (marked by the red rectangle) and a tapered region. The electric field distribution for the forward and backward propagation waves are shown in (a) and (b), respectively.
    The width of the linearly tapered region changes from $1$ to $0.5$. The field is only amplified in the forward direction in (a) but not reversed in (b).
    }
  \label{fig:FD_m1_dg_fb}
\end{figure}

In the structure shown in Fig.~\ref{fig:m1in_dg_fb}, the direction-dependent amplification for the even mode is accompanied by the generation of amplitudes in the odd mode. To provide a single mode response with directional amplification, one can use a passive reciprocal structure to filter out the odd mode.
An example is shown in Fig.~\ref{fig:FD_m1_dg_fb}, where we have used a tapered waveguide region as a modal filter.
The tapered region has the same dielectric constant as the waveguide. It has a length of $25$, and its width linearly changes from $1$ to $0.5$.
As is shown in Fig.~\ref{fig:FD_m1_dg_fb}a, in the forward direction, an input mode $\ket{1}$ is amplified by the gain-loss modulation region.
Then, since the generated mode $\ket{2}$ is not guided in the narrower part of the tapered region, it leaks out of the waveguide, leaving an amplified mode $\ket{1}$ at the output on the right side. In contrast, in the backward direction, input mode $\ket{1}$ propagates through the tapered region and the modulated region without any amplification, as is shown in Fig.~\ref{fig:FD_m1_dg_fb}b.
We note such a filter scheme is based on the modal profile rather than the frequency and hence can preserve the broad-band nature of the device, even in the case where the modulation frequency is small.

With directional gain available, it is straightforward to construct non-reciprocal optical isolation. One can connect the gain-loss modulation region with an absorption region, so that the wave has a net unit-transmission in one direction, while it is attenuated in the other direction.
The strength of optical isolation of such a device is tunable. One can tune the contrast ratio, defined by the ratio between the transmission coefficients in the two directions,  by adjusting the length of the gain-loss modulated region.

In the above demonstration, we have assumed a modulation frequency of $\Omega/\omega\approx 0.3$, and a modulation strength of $\delta\sigma=1$, which corresponds to a modulation strength in the imaginary part of permittivity of $\delta\varepsilon_i/\varepsilon= \delta\sigma/\omega\varepsilon\approx 0.1$.
In today's semiconductor laser technology, the achievable modulation frequency is a few tenth of gigahertz \cite{Nakahara2015}, corresponding to a smaller modulation frequency of $\Omega/\omega\approx 10^{-4}$.
The gain coefficient in these lasers typically reaches well over  $5\times 10^3\,{\textrm {cm}}^{-1}$ \cite{Ma2013}, corresponding to a large gain-loss modulation strength of $\delta\varepsilon_i / \varepsilon \geq 0.1 $.
Using the coupled mode theory formalism as discussed above, a non-reciprocal directional gain of $15$\,dB/mm can be achieved in a waveguide similar to what was shown in Fig.~\ref{fig:waveguide_modes}, assuming a realistic modulation frequency of 50\,GHz, and a modulation strength of $\delta\varepsilon_i/\varepsilon=10^{-3}$ \cite{Note1}.
We emphasize here that the method of gain-loss modulation  is directly integrable with standard semiconductor laser technology. It can be fabricated as a section of a diode laser waveguide, requiring no additional materials or photonic integration.

In summary, we have shown that dynamic gain-loss modulations in a dielectric waveguide structure can give rise to a direction-dependent $\mathcal{PT}$ phase transition that is thresholdless in the forward direction but with a non-zero threshold in the backward direction.
This effect can be used to achieve non-reciprocal directional gain that is broad-band, and the structure is directly integrable with standard semiconductor lasers. Our work points to a previously unexplored connection between $\mathcal{PT}$ symmetry and non-reciprocal physics.
Further exploration of this connection may offer new opportunities in studying novel non-Hermitian topological physics in dynamic and non-reciprocal systems \cite{Cerjan2018,Leykam2017,Zeuner2015,Malzard2015,Esaki2011}.

\begin{acknowledgments}
    This work is supported by  U. S. Air Force Office of Scientific Research (FA9550-16-1-0010, FA9550-17-1-0002).
\end{acknowledgments}


%

\end{document}